# Spreadsheet Auditing Software

*David Nixon, Mike O'Hara*
*Information Systems Institute, University of Salford, United Kingdom*

**ABSTRACT**

*It is now widely accepted that errors in spreadsheets are both common and potentially dangerous. Further research has taken place to investigate how frequently these errors occur, what impact they have, how the risk of spreadsheet errors can be reduced by following spreadsheet design guidelines and methodologies, and how effective auditing of a spreadsheet is in the detection of these errors. However, little research exists to establish the usefulness of software tools in the auditing of spreadsheets.*

*This paper documents and tests office software tools designed to assist in the audit of spreadsheets. The test was designed to identify the success of software tools in detecting different types of errors, to identify how the software tools assist the auditor and to determine the usefulness of the tools.*

## 1. INTRODUCTION

Despite much research into the causes of spreadsheet errors [Panko, 2000], and the ways of avoiding spreadsheet errors, these errors are still the rule rather than the exception. Initial error rates in spreadsheet development and debugging are consistent with those in traditional programming Error rates in 'live' spreadsheets are so high because the steps, followed in traditional programming to reduce these errors, are overlooked in spreadsheets. Errors can be classified, based on types of error, source of error and/or time of error [Rajalingham et al, 2000]. Whilst cell-by-cell team code inspection proves the most successful method of spreadsheet auditing, this still only produces an 80% success rate. Software Engineering concepts can be applied successfully to spreadsheet model development and auditing [Grossman, 2002]. Some spreadsheet development methodologies / guidelines are based upon Software Engineering principles and concepts [Knight et al, 2000]. Good spreadsheet design is essential in reducing spreadsheet errors [Raffensperger, 2001] and increasing readability and maintainability. An efficient use of resources will usually require some form of risk analysis [Butler, 2000] to decide to what depth a spreadsheet model should be audited. The conceptual difference between what a user sees, and what the computer sees, in a spreadsheet is the main reason that auditing spreadsheets is so difficult. Auditing is helped by using a visual approach to cell descriptions [Chen & Chan, 2000]. The visual auditing approach lends itself to the use of software auditing tools.

From the review of the academic research performed relating to spreadsheet errors, it was concluded that auditing of spreadsheets is incredibly difficult, particularly without the use of a visual auditing approach. In order to address this issue, software has been produced to assist in the audit of spreadsheets. These software tools tend to provide a visual approach to assist the user in auditing the spreadsheet. Most also tend to be aimed at the spreadsheet developer, often providing additional functions to assist in the development of spreadsheets as well as providing auditing functions. These tools range from those that merely assist in cell inspection and audit, to those that attempt to identify unique formulae and potential problem cells. To date, little, if any, research has been published that assesses the usefulness and capabilities of these software tools. This paper documents an investigation into software auditing tools and attempts to answer the following questions:

- Are software auditing tools for spreadsheets useful?
- How do software auditing tools assist the user in auditing spreadsheets?
- What errors are software auditing tools good/poor at identifying?

To answer these questions, four software auditing tools, along with the auditing functions built in to Excel, were tested against a specially designed spreadsheet that contained seeded errors. The software tools tested were:

- Excel's Built-In Auditing Functions - These are included as standard functions in Microsoft Excel, and for the purpose of the test were deemed to be primarily the functions available on the Excel auditing toolbar. http://support.microsoft.com/kb/289245

- The Spreadsheet Detective - A commercially available product that provides extensive auditing functions http://www.spreadsheetdetective.com/

- The Excel Auditor - A commercially available product that provides functions for both auditing and development of spreadsheets http://www.bygsoftware.com/auditor/auditor.htm

- The Operis Analysis Kit - A commercially available product that provides auditing and development functions for spreadsheets http://www.operis.com/oak.htm

- Spreadsheet Auditing for Customs and Excise (SpACE) - A tool in use at Customs and Excise to allow auditors to audit a business' VAT calculation spreadsheets. http://www.auditwaresystems.co.uk/product.asp?productID=11

## 2. THE SOFTWARE AUDITING TOOLS TEST

2.1 The Sample Errors Spreadsheet

In order to investigate different spreadsheet auditing software, a Sample Errors Spreadsheet was developed. This spreadsheet is loosely based upon a spreadsheet in use at the CWS, that is used to produce flash turnover reports in the case of an OLAS system failure (OLAS is the accounting software used by CWS Retail). The spreadsheet is split into three worksheets. The first of these is used to enter flash turnover data, the second sheet is used to hold the data downloaded the previous week from OLAS, and includes the rest of the data needed to produce the report held on the third worksheet. Once the data has been manually entered into the first worksheet, and the file loaded into the second worksheet, the report can be produced and distributed. Figure 1 shows the errors added to the spreadsheet:

Figure One – Seeded Errors

| Error Numbers | Error Type | Description |
|---|---|---|
| 1 to 4 | Qualitative | Formulas overwritten with values |
| 5 | Qualitative | Fixed value used when a named cell reference should be used |
| 6 | Qualitative | Presentation illogical |
| 7 | Qualitative | Incorrect Protection |
| 8 to 10 | Quantitative: Mechanical | Incorrect summation |
| 11 & 12 | Quantitative: Logical | A formula copied down a column looks at the wrong column |
| 13 | Quantitative: Logical | Incorrect Summation of a column including subtotals |
| 14 | Quantitative: Logical | Percentages added rather than calculated on a total cell |
| 15 | Quantitative: Logical | Percentages averaged rather than calculated on a total cell |
| 16 | Quantitative: Omission | A row missing from all the worksheets |
| 17 | Quantitative: Omission | A row missing from one of the worksheets, although it exists on the others |

## 2.2 Method Used to Test Software Tools

In order to maintain a level of consistency across the tests of spreadsheet tools, where possible, certain guidelines were followed. The tests each followed a three-stage procedure. The first stage involved the author becoming familiar with the software by examining it in an open session in order to gain a thorough understanding of the capabilities and limitations of the software, the way the software works and for whom the software was intended. This first stage allowed the author to investigate functions included in the software tool that were beyond the scope of the test errors in the Sample Errors Spreadsheet, to study any documentation provided with the software tool, and to become familiar with the software. The second stage involved the testing of the software tool against the Sample Errors Spreadsheet and the completion of the Test Results section of the Spreadsheet Auditing Software Test Form, including the allocation of a result to each error tested based upon the following criteria:

Figure Two – Pass/Fail Criteria

| PASSED | The software spotted the error with minimal intervention from the user, often highlighting the problem before the user examined the erroneous cell(s) on an individual basis. The software made the error much easier to spot by using visual display methods. |
|---|---|
| ALMOST-PASSED | The software hinted strongly at the possibility of the error and after a little investigation, the user, aided by the display tools in the software, was able to find the error. |
| ALMOST-FAILED | The software suggested the presence of an error but failed to identify the rogue cell(s). An investigation by the user found the offending error. |
| FAILED | The software failed to provide any suggestion that the error existed. |

The second stage tests were, with the exception of the Excel built in auditing functions test, completed by allowing the software to guide the author to identification of each error. The Excel built in auditing functions were, due to the limitations in the way they work and the fact that they are available for use alongside the other tools, tested on a cell-by-cell inspection basis. The final stage of the test involved the documentation of any additional features, problems and an overall impression of the software.

The exception to the aforementioned three-staged approach to testing the software was SpACE. Unfortunately it was not possible for the author to acquire a copy of SpACE for testing purposes without attending a training course in its use. As there was no training course scheduled before the submission date of this paper, it was impossible to test SpACE in the same three staged approach used in the tests of the other software tools. As a result, it was agreed that the second stage, the test against the Sample Errors Spreadsheet, would be done by an experienced user of SpACE, under the observation of the author, with the author awarding results based upon the observation of the test and completing the Spreadsheet Auditing Software Test Form after the test had taken place. Screenshots and reports were taken from each of the tests to allow future analysis of the results.

Due to the nature of the errors and the result criteria, along with the fact that the tests took place over a period of time, the results of the test could be classed as largely subjective. To attempt to reduce any subjective aspect of these results a normalisation stage took place once all of the tests had been completed. This involved examining each error in turn across the five software tools tested and where necessary, adjusting the results slightly to establish a level of consistency and to reduce the subjective elements of the results. Although a degree of subjectivity undoubtedly remains in the test results, because they are based upon questions of opinion such as "with this information in this format, how likely would an auditor be to identify this error?", as the expressed primary aims of the tests were to assess usefulness, methods and detection rates across error types, it was felt that a certain level of subjectivity in the test results could be tolerated, particularly as the results were to be drawn from five different software tools.

### 2.3 Excel's Built In Auditing functions

The first software to be tested against the Sample Errors Spreadsheet was Excel's own built in auditing functions. These built in functions were unique, in that, as they are included in Excel as standard, they were also used to supplement other software tools tested. As a general rule, the test utilised the built in functionality when testing other software tools only when the software had indicated potential errors within the cell or range of cells.

In the test, which was performed largely on a cell-by-cell inspection basis, the built in functions successfully highlighted the four 'formulas overwritten by values' errors, using both the view formula mode on the offending worksheets, and by showing that the cells in question did not contain any precedents. The functions failed to identify the use of a constant in a formula as a potential error, and had no options to identify patterns in labels on the report sheet so failed to identify the 'illogical presentation' error. As the built in functions required the user to remove worksheet protection before use, and makes no attempt to identify unprotected cells, the functions failed to find the 'incorrect protection' error. The built in functions had more success on the three mechanical errors, only struggling on the formula that involved a totalling of separate rows rather than a whole column. On the first three logical errors (domain errors) the functions clearly showed the constituents of the formulae, but this gave no suggestion that the constituents were incorrect. The built in functions fared slightly better on the final two logical errors (totalling and averaging of percentages) although it would still be possible for the user to overlook these errors, particularly if the user did not posses sufficient domain knowledge. The built in functions failed to indicate any possible errors in the final two errors in the test, both omission errors.

The major drawback of the built in auditing tools in Excel is the fact that in the majority of cases, the tools are used on an individual cell basis that means that the auditor is, in effect, doing a cell-by-cell inspection. Whilst this has proven successful in tests, and the auditing functions provide a more visual representation of the cell under investigation, a major reason for using a software auditing tool is to reduce the time spent on auditing spreadsheets by avoiding the time-consuming cell-by-cell method. The built in tools prove useful in the investigation of single cells and as a result can be used alongside other software auditing tools. The built in functions do not, however, provide the user with any guidance to identify potentially unsafe cells.

### 2.4 The Spreadsheet Detective

The Spreadsheet Detective is an Excel add-in and is produced on a shareware or licenceware basis by Southern Cross Software. The full registered version costs approximately £ 100. Unlike the test on the built in auditing functions in Excel, the test of The Spreadsheet Detective was not done on a cell-by-cell inspection basis.

There are two fundamental ways in which the software attempts to assist the user in the auditing of the spreadsheet. The first is the identification of formula schema, which should mean that the user has to check far fewer formula than under a cell-by-cell inspection method, as copied formula are identified as belonging to the same schema and therefore need not be checked. The second is the listing of potential problems such as references to non-numeric cells or unprotected schema in the footer report.

On the test itself, The Spreadsheet Detective performed particularly well, only really struggling on the errors that were particularly difficult to find. The annotate sheet function which provided an overlay to a worksheet with different types and colours of shading and text descriptions of formulae, meant that the first four errors concerning formula that were overwritten as values were obvious.

The footer report included a list of potential error cells including the 'constant in formula' error so this was also easy to find. The software failed to highlight the 'illogical presentation' error, although this could come to light when the adjacent cells are examined as they were deemed to be within a new formula schema as the order had changed. The 'incorrect protection' error was particularly easy to identify as unprotected cells were listed in the footer report despite the fact that worksheet protection had been disabled.

The Spreadsheet Detective achieved two 'almost-passed' and a 'passed' result for the three incorrect summation errors. It achieved the 'passed' result because of the automatic naming function which shows the formula in an English language type format, based upon the row and column headings, which meant that the error resulting from totalling the wrong rows stood out. The two logical domain errors, involving formulae looking at the wrong columns, were also much easier to find due to the automatic naming function. The error concerning the total formula that included both sub totals and detail lines was highlighted in the footer report under references to non numerics as it also included blank cells, so this was also easy to identify. A combination of the new schema shaded overlay, and the automatic naming function, meant that the totalling and averaging of percentages errors were relatively easy to identify. The lack of any un-referenced cells identification, or label pattern recognition, meant that the software did not highlight the final two omission errors.

The Spreadsheet Detective proved successful in the test, and has many functions that go well beyond those errors in the test spreadsheet. The two pronged approach of using an overlay to identify formula schema and types of cell, and having a summary report of potential errors, meant that the user was guided towards many of the errors by the software, the summary report proving particularly useful in identifying the more subtle errors such as unprotected ranges. The automatic naming of the constituents of formulae worked well in the errors spreadsheet, although this may prove less useful in a spreadsheet that is not as logically laid out with easy to identify row and column labels. However, this naming function remains one of the best features of the software and could even encourage developers to design spreadsheet models in a more logical way, to comply with these naming conventions.

## 2.5 The Excel Auditor

This software is produced by BYG Software. The Excel Auditor is an add-in for Excel and, despite its name, provides many functions outside the usual scope of auditing software.

The Excel Auditor provides two primary and two secondary auditing tools. The primary tools being the Audit Map, which provides a traditional audit map of a worksheet, and the Worksheet Cell Analyser, which documents the contents of a worksheet's cells. Both of these tools produce a report on a separate workbook. The secondary tools are the Cell Roots and Circular Reference Analyst functions.

On the test itself, The Excel Auditor performed poorly. Unlike the test of Excel's built in auditing functions, The Excel Auditor was not tested on a cell-by-cell inspection basis. This was because a primary aim of a supplementary auditing package should be to save the auditor the time that would be taken if a cell-by-cell inspection were necessary, and also to keep the test consistent with the other supplementary tools tested in this report. Unfortunately this meant that The Excel Auditor did not perform as well as the Excel functions due to the method of testing.

As a cell-by-cell inspection was not used with The Excel Auditor, the first four errors, concerning formulae overwritten with values, whilst clearly shown on the audit map, could easily be overlooked as they would require the use of the audit map alongside the original worksheet. In this case it could be argued that it is easier to move the cursor over each cell in the worksheet and watch the Excel input bar. As a result, The Excel Auditor only achieved an 'almost-failed' result for each of these errors. The Excel Auditor also achieved an 'almost-failed' result for the 'constant in formula' error as, whilst it clearly showed the constituents of the formula in the worksheet cell analyser, it failed to warn of any potential problems with this approach. The software failed to identify any problem with the 'illogical presentation' error and whilst the documentation function showed which worksheets had protection enabled, it could not do the same for individual cells.

On the three mechanical errors that related to incorrect summation, The Excel Auditor, whilst clearly documenting the formulae concerned, failed to suggest any problems with the formulae and therefore only achieved an 'almost failed' result for each of these errors. The same issues surfaced on the two logical/domain errors concerning 'incorrect formula' due to the wrong column being used' and the summation formula that included sub total rows as well as detail rows. Both of the incorrect percentage calculations suffered from the fact that they were buried away in the 64 page worksheet analysis report produced for the report sheet, although the average calculation was slightly easier to identify as an error. The final two omission errors both failed to be identified by The Excel Auditor, as there is no way of looking for label patterns or efficiently looking for cells with no dependents.

Perhaps the biggest problem with The Excel Auditor as an auditing tool is that is still requires cell-by-cell inspection to allow the auditor to confidently audit the spreadsheet. The software is also let down by the lack of visual aids to auditing, particularly by the fact that it functions on separate reports, meaning that the user is required to jump from report to spreadsheet whilst auditing the spreadsheet. The Excel Auditor is more likely to be of use as a documentation tool, rather than an auditing tool, as the documentation tools, although not covered in the test, are quite useful and easy to use.

2.6 The Operis Analysis Kit

This software is available from Operis Business Engineering Limited and is in the form of an Excel add-in. To perform spreadsheet auditing, Operis Analysis Kit adopts a two-pronged approach. The first allows the user to search the worksheet for particular types of cells that are more likely to be problematic, such as those with no dependents or those with formula containing hardwired constants. The second approach concentrates on graphically mapping the cells on a worksheet overlay to identify formula schema, referred to in Operis Analysis Kit as distinct formula, and types of cell, and to document the distinct formula and relevant statistics.

In the test itself, the four 'formulas overwritten with values' errors were easily identified using the option to overlay colour-coded cells to the worksheets. The search for hardwired constants option correctly identified the 'fixed value in a formula' error. The software did not, however, highlight any potential problem with the 'illogical presentation' error, although this could be spotted when the adjacent formula cells were examined as they were deemed to have distinct formula, so an 'almost-failed' result was recorded for this test. Operis Analysis Kit failed the 'incorrect protection' error, as it makes no attempt to identify which cells are left un-protected. The three mechanical errors all achieved an 'almost-passed' result as each was identified as a distinct formula and the contents of these formula were clearly documented.

The two logical/domain errors achieved 'almost-failed' results, as although the formula was marked as distinct in each case, and the formulae were clearly documented, there was no suggestion that the formulae were incorrect. The 'incorrect summation of a column including sub totals' error was easily identified using the search option to find references to blank cells. The two incorrect percentages tests produced contrasting results. On the first of these tests, which involved the totalling of percentages, Operis Analysis Kit achieved a 'failed' result as the formula was not identified as distinct as it matched the formula in the cell to the left, whereas on the second test, that of the averaging of percentages,

Operis Analysis Kit achieved a 'passed' result as this was correctly identified as a distinct formula and upon examination of the cell documentation was obvious. The two omission errors also produced contrasting results; with the first error achieving a 'failed' result, as there were no options to identify label patterns, whereas the second error achieved a 'passed' result, as using the search for unreferenced cells on the file worksheet correctly highlighted the problem.

Operis Analysis Kit as a whole is very easy to use thanks to its operation via a simple additional menu in Excel. It is deceptively powerful, particularly in the search options, and managed to at least hint at all of the more 'findable' errors with the exception of its inability to identify unprotected cells.

2.7 Spreadsheet Auditing for Customs and Excise (SpACE)

SpACE was developed in-house by HM Customs and Excise for use by VAT inspectors in auditing client's spreadsheets, and is now available to the public. SpACE works by using a combination of search facilities, overlaid mapping options and the identification of unique formula, to attempt to highlight potential errors in a spreadsheet.

The test of SpACE was done slightly differently to the other software tool tests as the test was performed by Ray Butler, an employee of HM Customs and Excise and a regular user of the SpACE software, under the observation of the author. Unlike the author, Ray did not know the location of the seeded errors in the sample errors spreadsheet before the test. SpACE proved exceptionally good on the test itself, recording at least an 'almost-passed' result on each error test.

The software easily passed the first four 'formula replaced with values' tests using either the colour coded overlay, or the identification of numeric cells with no dependants option. SpACE achieved an 'almost-passed' result in the 'fixed value in formula' error as the software allows the user to search for fixed values, but the user has to enter the value to be found. SpACE achieved an 'almost-passed' result on the 'illogical presentation' error as it allowed the user to search for particular text strings and apply colour coding which, along with the indication that the adjacent formula cells do not follow the same pattern as the other should highlight the problem.

SpACE successfully found the 'incorrect protection' error as, despite automatically unprotecting the worksheets, the software is still able to identify the individual cells that have been marked as unprotected. In line with most of the other tools tested, SpACE correctly identified the three mechanical errors (incorrect summation errors) as unique formula yet the lack of a visual report means that this error could be overlooked by the user. The two logical/domain errors both achieved an 'almost-passed' result, thanks largely to the SpACE option of attempting to link user specified root and bottom-line cells.

When the software failed to establish a link between these two cells, further investigation revealed the errors. The 'incorrect summation of column containing subtotals' error was easily found by SpACE as it was highlighted as a unique formula, which upon further investigation shows the error to be obvious. Both of the erroneous percentage calculations were highlighted by SpACE in a way that was obvious enough to earn a 'passed' result. SpACE was the only software tool that had any success with the first omission error, which involved a row missing from every sheet in the workbook. It achieved an 'almost-passed' result by a combination of a schema identification method that allows 'blocks' of formulae to be grouped together, and the option to highlight text strings with different coloured backgrounds. The final omission error, achieved a 'passed' result on SpACE as the option to identify cells with no dependencies highlighted the offending row on the file worksheet.

SpACE is a 'tried and tested' software tool and has obviously been in use and subject to improvement for some time. In addition to the functions covered in the test, SpACE includes more in-depth auditing tools such as the ability to check lists of data for duplicates and attempt to identify the numbers that make up a particular total. The only major function missing from SpACE, that was present in any of the other software tools tested, appears to be some form of English language type formula description tool similar to that found in The Spreadsheet Detective.

## 3. TEST RESULTS

### 3.1 Results by Software Tools

As can be seen in the Figures Three and Four, SpACE, The Spreadsheet Detective and Operis Analysis kit show test results of 43 (84%), 41 (80%) and 33 (65%). It is possibly significant however that the SpACE results were assisted by the fact that the test was carried out by an experienced SpACE user (although the results were allocated by the author), rather than by the author, who was new to all of the other software before the tests began. It should also be noted that even on a cell-by-cell inspection basis, the built in Excel functions only score 24 (47%), which suggests that all is not well and that the need to go beyond the built in auditing functions exists.

Operis Analysis Kit, which is probably the easiest tool to use and scored a respectable 33 (65%). The Excel Auditor only scored 14 (27%), although this was largely down to the method of testing as it needs to be used as a tool to assist in a cell-by-cell inspection, whereas it was tested on the same basis as the other three add-in packages.

Figure Three – Summary Test Results

|  | Built-In Excel Functions | The Spreadsheet Detective | The Excel Auditor | Operis Analysis Kit | SpACE |
|---|---|---|---|---|---|
| Passed | 0 | 12 | 0 | 8 | 9 |
| Almost Passed | 8 | 2 | 1 | 3 | 8 |
| Almost Failed | 4 | 1 | 12 | 3 | 0 |
| Failed | 5 | 2 | 4 | 3 | 0 |
| Total | 17 | 17 | 17 | 17 | 17 |

During the test, it became apparent that certain functions were more useful than others, both in the Sample Error Spreadsheet test itself, and in the familiarisation stage of the test where other functions were investigated.

Possibly the most essential ability for a software tool to possess is the ability to recognise formula schema. It is this ability that lifts the software from being a tool to assist in a cell-by-cell inspection, to a tool that means a time-consuming cell-by-cell inspection can be avoided.

A second function that could be classed as essential is the ability to provide the user with a visual overlay on the worksheet, identifying different schema/different types of cells etc. Falling into the 'very useful' category are functions such as supplementary potential error cells reports, the option to search for particular cells that are prone to errors, such as constants in formulae or un-referenced cells, the ability to identify unprotected cells, and the use of English language formula descriptions based upon column and row labels.

It is in these functions that SpACE and The Spreadsheet Detective excel, and that enabled them to score so highly in the tests. Other features such as documentation and development tools, whilst not contributing to the test results due to the nature of the Sample Errors Spreadsheet, could certainly be useful, depending upon who the user of the software is and the role they perform i.e. developer/auditor/end-user.

Figure Four – Detailed Test Results

| Type | Errors Description | Number | Built-In Excel Functions * | The Spreadsheet Detective | The Excel Auditor | Operis Analysis Kit | SpACE | Total | Average Score | Average Score for type of error | Average Score excl Excel built In |
|---|---|---|---|---|---|---|---|---|---|---|---|
| Qualitative | Formula overwritten with values | 1 | 3 | 3 | 1 | 3 | 3 | 13 | 2.6 | | 2.5 |
| | Formula overwritten with values | 2 | 3 | 3 | 1 | 3 | 3 | 13 | 2.6 | | 2.5 |
| | Formula overwritten with values | 3 | 3 | 3 | 1 | 3 | 3 | 13 | 2.6 | | 2.5 |
| | Formula overwritten with values | 4 | 3 | 3 | 1 | 3 | 3 | 13 | 2.6 | | 2.5 |
| | Fixed value used within formula | 5 | 0 | 3 | 1 | 3 | 2 | 9 | 1.8 | | 2.3 |
| | Presentation Illogical | 6 | 0 | 1 | 0 | 1 | 2 | 4 | 0.8 | | 1.0 |
| | Incorrect Protection | 7 | 0 | 3 | 0 | 0 | 3 | 6 | 1.2 | 2.0 | 1.5 |
| Quantitative – Mechanical | Incorrect Summation | 8 | 2 | 2 | 1 | 2 | 2 | 9 | 1.8 | | 1.8 |
| | Incorrect Summation | 9 | 1 | 3 | 1 | 2 | 2 | 9 | 1.8 | | 2.0 |
| | Incorrect Summation | 10 | 2 | 2 | 1 | 2 | 2 | 9 | 1.8 | 1.8 | 1.8 |
| – Logical | Copied formula looks at wrong column | 11 | 1 | 3 | 1 | 1 | 2 | 8 | 1.6 | | 1.8 |
| | Copied formula looks at wrong column | 12 | 1 | 3 | 1 | 1 | 2 | 8 | 1.6 | | 1.8 |
| | Incorrect Summation of a column (including subtotals) | 13 | 1 | 3 | 1 | 3 | 3 | 11 | 2.2 | | 2.5 |
| | Percentages added rather than calculated | 14 | 2 | 3 | 1 | 0 | 3 | 9 | 1.8 | | 1.8 |
| | Percentages averaged rather than calculated | 15 | 2 | 3 | 2 | 3 | 3 | 13 | 2.6 | 2.0 | 2.8 |
| – Omission | A row missing from all the worksheets | 16 | 0 | 0 | 0 | 0 | 2 | 2 | 0.4 | | 0.5 |
| | A row missing from one of the worksheets although it exists on others | 17 | 0 | 0 | 0 | 3 | 3 | 6 | 1.2 | 0.8 | 1.5 |
| Total Score | | | 24 | 41 | 14 | 33 | 43 | 155 | 31.0 | 1.8 | 32.8 |
| Percent of Possible | | | 47% | 80% | 27% | 65% | 84% | | 61% | | |

*Test done on a cell-by-cell basis

Scoring System
Passed — 3
Almost-Passed — 2
Almost-Failed — 1
Failed — 0

3.2 Results by Type of Error

When the results of each of the software tests are combined, it is possible to gain an overview of the levels of success that were achieved in relation to each type of error. From the results of the test, it is obvious that the software tools tested performed to a reasonable standard on all but the omission errors, for which the average result was 0.8, far below the average results reported for the other three error types. This is not altogether surprising, as these errors are the type that an auditor is more likely to identify with the human eye than with a software tool.

Those tools that highlighted these omission errors relied upon non-referenced cells elsewhere in the spreadsheet model, and identifying label/formula patterns. The other three error types all achieved good results in the tests, although the mechanical errors did not quite match the qualitative and logical errors. Perhaps the most surprising result was the success of the qualitative errors, although these were improved slightly by having a particularly easy to spot (yet quite common) error in four different places. Considering that a definition of a qualitative error is that it does not (yet) have an impact on the bottom line figures, it was anticipated that these would be particularly difficult to find. The results on all but the omission errors do suggest however that software tools can have a positive effect on software auditing detection and efficiency.

3.3 Results by Individual Errors

Figure Five - Average Results for Individual Errors

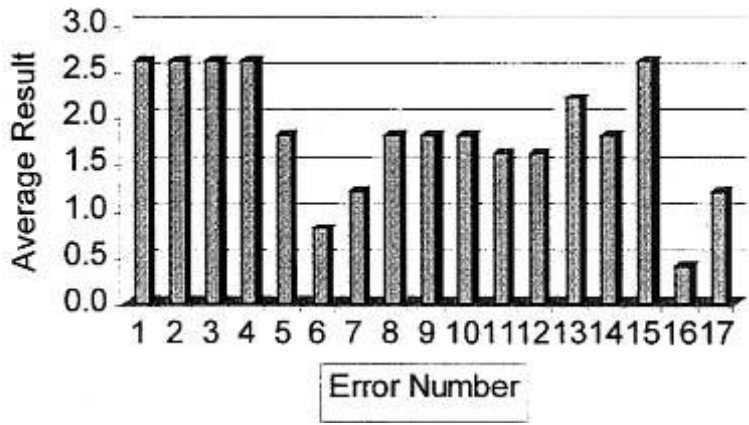

Formulas overwritten with values (errors 1 to 4) On these errors, the results averaged 2.6, a particularly successful result that illustrated how easy these errors can be to detect. Only The Excel Auditor of the add-in software tools tested, failed to achieve a 'passed' result. The success in finding these errors was largely down to the colour-coded overlays that could be applied to the worksheet making these errors stand out.

Fixed value used when a named cell reference should be used (error 5) This error achieved a respectable result of 1.8, which when the cell-by-cell inspection method used in the built in Excel functions test was removed, increased to 2.3. Both The Spreadsheet Detective and Operis Analysis Kit explicitly highlighted this error via an error report and a search option respectively. SpACE was close behind with a search option but this option needed the value to be specified by the user.

Presentation Illogical (error 6) The presentation illogical error proved very difficult to detect, with SpACE getting the closest by allowing certain text strings to be formatted with a different background colour meaning that text patterns were easier for the user to identify. Other than this, Operis Analysis Kit and The Spreadsheet Detective both partially identified this error by virtue of the fact that the adjacent cells had new formula schema.

Incorrect Presentation (error 7) This error surprisingly proved quite difficult to identify in the tests. However both The Spreadsheet Detective and SpACE correctly identified the unprotected range of cells despite having previously disabled the worksheet protection. The Excel Auditor, Operis Analysis Kit and the built in Excel functions all failed to identify the erroneous cells, although The Excel Auditor did indicate which worksheets were protected.

Incorrect Summation (errors 8 to 10) These errors received a reasonable average result of 1.8. The more successful methods of finding these errors were the graphical overlays, which showed new formula schema and/or highlighted the cells being totalled. The exceptional result on these errors was by The Spreadsheet Detective highlighting the incorrect row error (error 9) by using the labels to indicate, in English, that the values being totalled were not from the correct rows.

A formula copied down a column looks at the wrong column (errors 11 and 12) These errors were designed to be difficult to detect and the result of 1.6 for each was higher than expected. Although each of the software tools were capable of highlighting the formula and displaying the formula in a readable style, the nature of this error is such that even when the formula is displayed clearly, the error is not always obvious without some domain knowledge. The only software tool to achieve a 'passed' result in this test was The Spreadsheet Detective. This was largely down to the English descriptions. SpACE achieved on 'almost-passed' thanks to its failure to establish a link between the source data and the bottom line cell, which prompted further investigation that subsequently detected the error.

Incorrect Summation of a column including subtotals (error 13) Whilst a visual overlay of this formula showed which cells were included in the formula, the nature of the error meant that this error could still be overlooked. 'Passed' results were achieved by The Spreadsheet Detective, Operis Analysis Kit and SpACE by highlighting the cell as a prospective error due to the fact that it contained blank cells.

Percentages added rather than calculated on a total cell (error 14) With a graphical representation of the formula in question this error is relatively easy to detect. The Spreadsheet Detective and SpACE both achieved a 'passed' result on this error by virtue of identifying a unique formula which when viewed with the built in Excel functions became obvious. Operis Analysis Kit, however, failed this test as the total cell to the left was the same logical formula and so the cell was not highlighted as a distinct formula.

Percentages averaged rather than calculated on a total cell (error 15) Although superficially similar to the previous error, this error proved to be much easier to detect. The fact that it was an average of percentages made the error easier to detect even on a formula report. This error was also highlighted as a distinct formula in Operis Analysis Kit meaning that this achieved a 'passed' result on this test, whereas a 'failed' result was achieved by Operis Analysis Kit on the previous error.

A row missing from all of the worksheets (error 16) The omission errors were deliberately included as a difficult test for the software tools. The first was designed to provide an error that was expected to be much more likely to be detected by the human eye than a software tool. The fact that the row in question was removed from all of the worksheets means that the software would have to have some way of identifying label patterns to identify the absence of the row. Only SpACE came close to identifying the absence of the row, thanks to a combination of text background colour coding and the ability to recognise unique formula across 'blocks' of formulae.

A row missingfrom one of the worksheets although it exists on the others (error 17) This omission error was slightly easier to find than the previous error as the row of data deleted from the Report worksheet still existed on the File worksheet. Both SpACE and Operis Analysis Kit achieved a 'passed' result on this test due to the option to highlight cells with no dependents.

## 4. TEST CONCLUSIONS

The test has successfully provided the following answers to the questions identified at the outset of the test.

Are software-auditing tools for spreadsheets useful? The tests proved conclusively that spreadsheet auditing software tools can be useful. The more successful software achieved detection rates of over 80%. However it must be remembered that software tools do not detect and correct errors, they merely point the user in the right direction.

How do software-auditing tools assist the user in auditing spreadsheets? The software tools used various different methods of identifying potential errors, however the most successful used a combination of the following three features:

- Schema Identification
- Visual Methods
- Search for/Report on potential error cells

The software tools tested tended to be designed as add-ins to Excel, and as such could be used alongside the built in Excel Auditing functions.

What errors are software-auditing tools good/poor at identifying? Software tools proved to be particularly good at detecting qualitative, quantitative logical and quantitative mechanical errors in the test. Software tools proved somewhat less successful at detecting quantitative omission errors. Given the nature of these errors, this was not a surprising result.

## 5. OVERALL CONCLUSIONS

Spreadsheet errors are still a major problem. Research has repeatedly found that rates of spreadsheet errors are in line with those in more traditional programming techniques. In the author's professional experience as a developer of a number of spreadsheet models, spreadsheets are not allocated as much time and resources as a traditional programming language would be for testing and auditing purposes. Cell-by-cell inspection is largely unheard of, and even dummy data testing can be limited. These experiences are confirmed as commonplace by the academic research published, thus confirming that the author's experiences are not unique.

Research has taken place, which has found that spreadsheet auditing is easier using a visual approach. A visual approach lends itself to software tools. However to date there has been no published research into the usefulness of and methods used in, spreadsheet auditing software tools. This paper provides an initial investigation into software tools. The evidence shows that software tools are definitely useful in the detection of software errors, although they perform poorly on omission errors. The software tools potentially have three methods of assisting in the audit of a spreadsheet. The first is by applying a visual approach to auditing to the user, and the second is to identify cells that potentially contain errors. The third method available to software tools, and potentially the most powerful, is the identification of formula schema, which allows the user to adopt a 'focused cell inspection' approach to spreadsheet auditing, rather than a traditional cell-by-cell inspection approach.

## 6. AREAS FOR FURTHER STUDY

Extensive research into spreadsheet errors has taken place. This research has covered identifying and categorising errors, quantifying errors, methods of designing spreadsheets to avoid errors and error detection. There is however, very little research into the use of software tools to detect errors in spreadsheets. This paper attempts to redress this by testing a series of software tools against a spreadsheet with seeded errors. Whilst the tests proved that spreadsheet auditing software is useful, certain limitations in the software suggest that further study in this field would be beneficial. The nature of the test meant that the results were largely subjective. The test was primarily carried out by the author, who also designed the sample errors spreadsheet. This meant that the author had to speculate how much user interaction would be needed to identify an error and award a result accordingly. A more objective test would be to have the test carried out by a group of people who did not know the location of the errors before the tests. Another limitation of the test was that the sample errors spreadsheet was based primarily on a financial reporting model. Spreadsheets are used for many different applications other than just financial applications. It would be useful, therefore, to perform the tests on a number of spreadsheets from different fields, such as a results analysis spreadsheet or a 'what if' based spreadsheet.